# Connectivity and Cost Trade-offs in Multihop Wireless Networks

May Lim[1,2,*], Dan Braha[1,3] Sanith Wijesinghe[1], Stephenson Tucker,[4] and Yaneer Bar-Yam[1]

[1]New England Complex Systems Institute
[2]Brandeis University
[3]University of Massachusetts Dartmouth
[4]Sandia National Laboratory

## Abstract

Ad-hoc wireless networks are of increasing importance in communication and are frequently constrained by energy use. Here we propose a distributed, non-hierarchical adaptive method using preferential detachment for adjusting node transmission power to reduce overall power consumption without violating network load limitations. We derive a cost and path length trade-off diagram that establishes the bounds of effectiveness of the adaptive strategy and compare it with uniform node transmission strategy for several node topologies. We achieve cost savings as high as 90% for specific topologies.

PACS numbers: 89.75.Hc (Networks and genealogical trees), 84.60.Bk (Performance characteristics of energy conversion systems; figure of merit), 84.40.Ua (Telecommunications: signal transmission and processing; communication satellites), 89.75.Fb (Structures and organization in complex systems)

*M.L. is on leave from the National Institute of Physics, University of the Philippines.

# I. Introduction

In multihop wireless networks, messages may traverse multiple wireless links in order to reach a destination [1]. Though still faced with technical design challenges [2], significant interest has been generated for applications in mobile phone communications [3], wireless broadband [4], and distributed sensor networks [5]. Ongoing component miniaturization promises to lead to a variety of uses for dynamic, spontaneous or rapid deployment of networks for a variety of physical contexts and uses [6]. While actual costs, sizes, and protocols may vary; these networks are physically a set of geographically distributed transmitters and receivers linked to one another based on the strength of a local power source. Given the variation in robustness to failure and attack [7, 8], navigability [9], jamming [10], and speed of information propagation [11] found so far in random, small world [12], and scale-free network topologies [13], it is of great technological and economic interest to study the network characteristics of different topologies and overall network costs.

The wireless nature of the network generally implies a lack of continuous (wired) power. The need for batteries implies a key constraint in wireless networks is the cost of the localized power required to establish links, with a usage that grows as the square of the transmission distance. Prior studies [1] have investigated strategies to control node location and thus network topology coupled to varying node transmission power. While it is clear that node topology greatly influences the overall network routing capacity, delay time and robustness, node spatial distribution is often constrained, e.g. by landscape topography or the means by which nodes are distributed, and cannot always be independently controllable. As such, widely-known methods of preferential attachment [13] and link rewiring [12] are not readily applicable. Instead a simple preferential detachment algorithm that minimizes the number of links in high-density areas while maintaining links in low-density areas is proposed. In this paper we first analyze uniform node radius networks for different nodal configurations, and second describe an adaptive algorithm that reduces total power consumption while respecting constraints of multihop path length.

**II. Network Model**

We consider $N$ nodes distributed in two-dimensional space represented by a square $s$ x $s$ grid where the $i$th node broadcasts signal packets over a radius $r_i$ from its location, $x_i$. If another node $j$ is located within $r_i$, a link is established from $i$ to $j$ with a distance of one hop from $i$ to $j$ ($d_{ij} = 1$). Unlike wired networks, a link from $i$ to $j$ does not imply a link from $j$ to $i$ due to possible differences in node transmission radii. Upon receiving a packet, a node may retransmit it in accordance with its routing protocol up to a maximum of $h_{max}$ hops. We assume the existence of an efficient protocol that satisfies a shortest path routing metric [14, 15]. Furthermore, all nodes are functionally identical and independent (non-hierarchical) controlling only their own transmission power and have no computational overhead (e.g. from route optimization).

The choice for the prespecified parameter $h_{max}$, can be made to achieve tradeoffs in transmission: a low $h_{max}$ would require high coverage radii resulting in overall high power consumption; while a high $h_{max}$, though having lower node coverage radii and consequent cost, has a higher network load by virtue of having more hops which increases the risk of network congestion. Thus, each network topology has an associated tradeoff diagram between path length and cost as a function of $h_{max}$.

We characterize network structures that result from a prespecified value of $h_{max}$ using (1) the reachable pairs fraction $R = n / N(N-1)$, where $n$ is the number of distinct ordered pairs $(i, j)$ such that $i$ can transmit to $j$ through less than $h_{max}$ links; (2) the average path length $L = \Sigma\, d_{ij}/n$ of all reachable pairs; and (3) a normalized network cost $C = (\Sigma\, r_i^2)/C_o$, where $C_o = r_o^2$ is the transmission power required for a node at the center of the grid to broadcast over the entire area in one hop ($r_o$ is half the grid diagonal length = $s\sqrt{2}/2$ and $d_{oi} = 1$ for all $i$).

Uniform radius networks have previously been studied as a percolation problem in a two-dimensional random lattice in which bonds are determined by the distance between sites [16]. In such a percolation problem, $N$ random sites are randomly located in an $s$ x $s$ square grid, and the overall scale is arbitrary. When the distance between the two sites is less than $r_s$, a bond is formed between them. At the critical (percolation) radius $r_s = r_{sp}$, one can find a series of links that traverses the space in each linear dimension, going either up-down or left-right. Monte Carlo simulation and analytic approximations have shown that $r_{sp} = (1.06 \pm 0.03)\, 2s/\sqrt{(\pi N)}$ [16]. For our uniform radius network, (Fig. 1b, $h_{max} = 20$ hops), we expect that the average path length $L$ reaches its peak value, $L(r_p \approx 50) \approx 10.5$ hops, near the critical radius of percolation at the same time that the reachable pairs fraction $R$ changes most rapidly. In the absence of an $h_{max}$ limit, $r_{sp}$ gives us a lower bound on the node transmission radius for boundary-to-boundary connectivity, albeit in the presence of "dead spots" indicating isolated clusters ($R < 1$). Imposing increasingly smaller $h_{max}$ would necessarily increase all limiting radii values, thus $r_{sp}$ is a strict lower bound. For the node distribution in Fig. 1a where $N = 256$ and $s = 600$, we calculate $r_{sp} = 45 \pm 1.3$, which is consistent with Fig. 1b. While percolation is concerned with widespread communication, our concern is complete communication. The relationship between the percolation problem and our multihop network ends with the determination of the lower bound: our interest is in finding the minimum radius $r_{min} > r_p$ with full nodal coverage ($R = 1$) which is necessarily a higher value ($r_{min} = 60$ in Fig. 1a). $r_{min}$ depends on the value of $h_{max}$.

Figure 1a shows the lowest power uniform network topology formed by $N = 256$ nodes randomly distributed over a 600 x 600 space with $r_i = r = 60$ for all nodes $i$, with $h_{max} = 20$ hops. The normalized network cost is $C = Nr^2/C_o = 5.12$. For this case of uniform node transmission radii the links are symmetric. The figure reveals two significant weaknesses of uniform radius networks given random node placement: (1) There are regions of high node densities where there are many linked nearby neighbors ($r >$ nodal distances) that are regions of unnecessary power consumption, and also potentially flashpoints for network congestion; and (2) There are regions of nodes in low density areas where reducing the power or node failure would lead to network fragmentation.

From these observations, it seems that reducing node transmission radius ($r_i$) in high node density locations would yield significant cost savings. However, setting the power based upon local density can be shown to be generally ineffective. Power reduction that occurs in regions of high density is often compensated for by unnecessary increases in power consumption for nodes outside that region when a single functional form is used to adjust power for all nodes. This can be seen through direct analysis of a three node network using, e.g. $r_i = r_{max}/\rho_{local}$ [17]. More generally, setting the radius based upon local density is ineffective because the density is not isotropic, so that nodes that are near the edges of clusters defeat optimization by simple algorithms. Indeed, simulations show that a variety of density dependent rules do not improve significantly on uniform radius protocols [17]. We are thus motivated to find an adaptive method that overcomes this limitation while achieving power reduction in dense clusters.

We consider adaptive adjustment of node radii using an algorithm that reduces a radius of nodes until a minimum criterion for effective communication is achieved. In addition to complete network connectivity, we further restrict the networks to have a prespecified maximum number of allowed hops $h_{max}$. The adaptive process begins with establishing the best uniform radius network by setting the uniform radius to some maximum value (typically of the order of $s$) and performing synchronous radii reduction. When the network connectivity breaks ($R < 1$) we incrementally increase the radius so that the network is fully connected. The adjustment of individual node radii then occurs as an asynchronous update of each node (according to a pseudo random permutation of the nodes) according to the following protocol:

1. Node $i$ broadcasts a signal to all nodes and requests acknowledgment of receipt. Nodes receiving the signal respond to the initial request. Signal retransmission is allowed until $h_{max}$ is reached.

2. Node $i$ decrements its radius by a fixed amount ($r_i' = r_i - r_d$).

3. Node $i$ resends an acknowledgment request packet to all nodes under the same $h_{max}$ constraint. All receiving nodes respond and allow retransmission until $h_{max}$ is reached.

4. If node $i$ receives the same number of replies, it updates $r_i = r_i'$. Otherwise, the node locks its transmission power to $r_i$ and no longer performs the above protocol in subsequent iterations.

The cycle repeats until all nodes have locked their power values (the number of such update cycles is bounded by $r_{init}/r_d$, where $r_{init}$ takes the value of the initial uniform radius. The nature of the algorithm ensures that the overall normalized network cost $C = \Sigma\ r_i^2/C_o$ is equal or better than that of a uniform node radii algorithm under the condition that $L$ cannot exceed $h_{max}$. Moreover, it is guaranteed that no node can reduce its transmission power without violating this condition.

We measure the efficacy of an adaptive network with respect to a uniform radius network using eight different nodal topologies for 256 nodes (Fig. 2, coordinate origin (0, 0) is the bottom-left corner of each panel) namely: A) random; B) random in three 200 x 200 clusters centered at x-y coordinates (100, 100), (300, 400), (500, 200) with 50 nodes per cluster, and the remaining nodes randomly distributed over the 600 x 600 grid; C) 60% within a 200-radius central cluster with coordinates ($\rho \cos\theta$, $\rho \sin\theta$), where $\rho$ is randomly generated in the range (0, 200) and $\theta$ in the range of (0, $2\pi$), 40% randomly distributed; D) star configuration (five 200 x 200 randomly-distributed clusters centered at x-y coordinates (100, 100), (100, 500), (500, 500), (500, 100), and (300, 300) with 50 nodes per cluster except the central cluster with 56 nodes; E) uniform lattice; F) radial with coordinates ($k \cos[2\pi k(k+1)/96] + 300$ m, $k \sin[2\pi k(k+1)/96] + 300$) and $k$ is an integer from 1 to 256; G) distributed along preset lines; and H) random walk starting at the center. These configurations were chosen to represent a variety of geographical or nodal deployment constraints.

### III. Results and Discussion

Figure 3 compares the network average path length ($L$) and cost ($C$) for the adaptive (circles) and uniform radii (square) methods for each node topology for varying $h_{max}$ (5 to 30 hops in 5-hop increments). In general, the adaptive method provides significant cost savings given the same $h_{max}$ over the uniform radii method. More specifically, we observe the following trends:

a) For larger values of $h_{max}$, the largest nearest neighbor distance sets the minimum value of $r$. Figure 3c shows that for $h_{max} > 5$, the uniform radii method reached its limit due to the presence of a single relatively isolated node (Fig. 2c, bottom center). In general, $L$ increases with $h_{max}$.

b) The uniform and adaptive solutions converge to the same value for large values of $h_{max}$ if and only if a constant nearest neighbor distance exists. In Fig. 3e, convergence is achieved at $h_{max} = 30$ hops: the optimum radius $r_{min}$ is the minimum node-to-node distance and 30 hops exactly cover the distance from corner-to-corner in the uniform grid. It is worth noting that the line topology (Fig. 2g) would have belonged to this category had all the lines been joined together. In practical situations, such a break in the line may have been caused by a few nonfunctional nodes at critical junctions and highlights the strength of an adaptive method that allocates increased power output only at the boundary nodes.

c) In regions of high density, the adaptive method significantly reduces node power output when subject to the $h_{max}$ constraint. The significant power savings can alternatively be used to reduce L for the same total power consumption (Fig 3).

In high speed transmissions with minimal latency, we may take $h_{max}$ to represent packet lifetime which is proportional to the number of live packets at any given time. Taking $h_{max}$ as a proxy for network load, we compare the performance of an adaptive algorithm against that of a uniform algorithm using two metrics: a) relative cost $C_{ratio} = C_{adaptive} / C_{uniform}$; and b) length factor $L_{ratio} = L_{adaptive}/L_{uniform}$. Figure 4 shows the tradeoff between $C_{ratio}$ and $L_{ratio}$ for different topologies and $h_{max}$ values. Topology-dependent effects are particularly evident for: a) topology C where increasing $h_{max}$ results in a marginal improvement of cost but results in a significant length factor change; and b) topology E where the uniform grid creates unique stepwise relationships between $h_{max}$ and $r$. In all other cases where a smooth distribution of the nearest neighbor distances exists, we obtain $C_{\text{ratio}} \sim 0.2$ which translates to about 80% cost savings at the expense of a two- or three-fold length factor change under the same network load ($h_{\max}$). In some cases the power savings exceed 90%.

**IV. Conclusions**

We have shown that an adaptive node radius distribution strategy given fixed node locations provides an average of 80% and as much as 90% cost savings over a uniform node radius allocation. This power savings is achieved with a fixed upper bound on the path length. The average number of hops increases by 2-3 fold. If communication path-length constrains the network use, the adaptive algorithm can be used to reduce the path length without increasing power consumption.

**Acknowledgments**

This work was supported in part by Sandia National Laboratories under US DoE Contract DE-AC04-94AL85000.

[17] M. Lim *et al.*, (unpublished).

**Captions to Figures**

Fig. 1. **a**, Network of 256 nodes (circles) in a 600 x 600 grid with uniform transmission radius $r = 60$. Links are shown where inter-node distance is one hop ($d_{ij} = 1$). **b**, Variation of the average path length $L(r)$ and reachable pairs $R(r)$ for the network in (a), $h_{max} = 20$ hops.

Fig. 2. Test nodal distributions: **a,** random; **b,** random in three clusters; **c,** 60% within a 200-radius central cluster, 40% randomly distributed; **d,** star configuration; **e,** uniform lattice; **f,** radial; **g,** lines; and **h,** random walk starting at the center.

Fig. 3. Panels correspond to node distributions (Fig 2) for adaptive (circles) and uniform radii (squares) methods. Points indicate the minimum possible cost $C$ and average path length $L$ given $h_{max} = 5, 10, \ldots , 30$. In general, $L$ increases with $h_{max}$.

Figure 4. Power cost gain $C_{ratio} = C_{adaptive}/C_{uniform}$ and path length change $L_{ratio} = L_{adaptive}/L_{uniform}$ tradeoffs in adopting an adaptive over the over a uniform radius method.

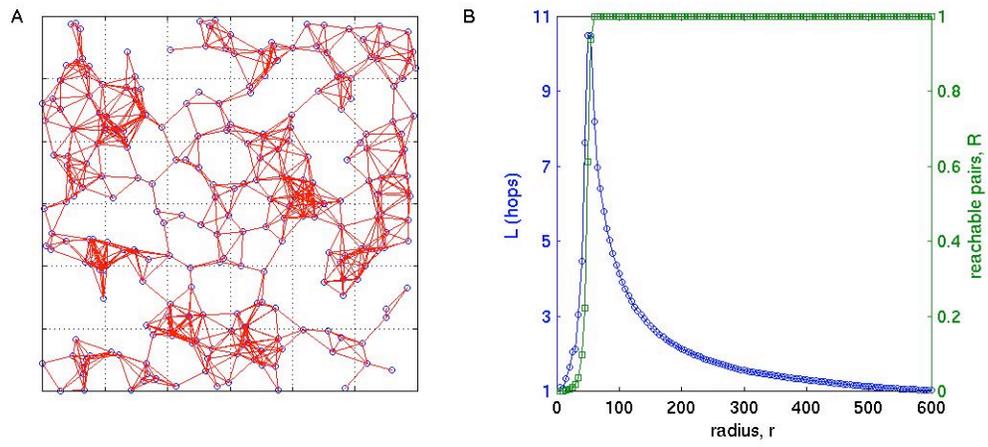

Figure 1.

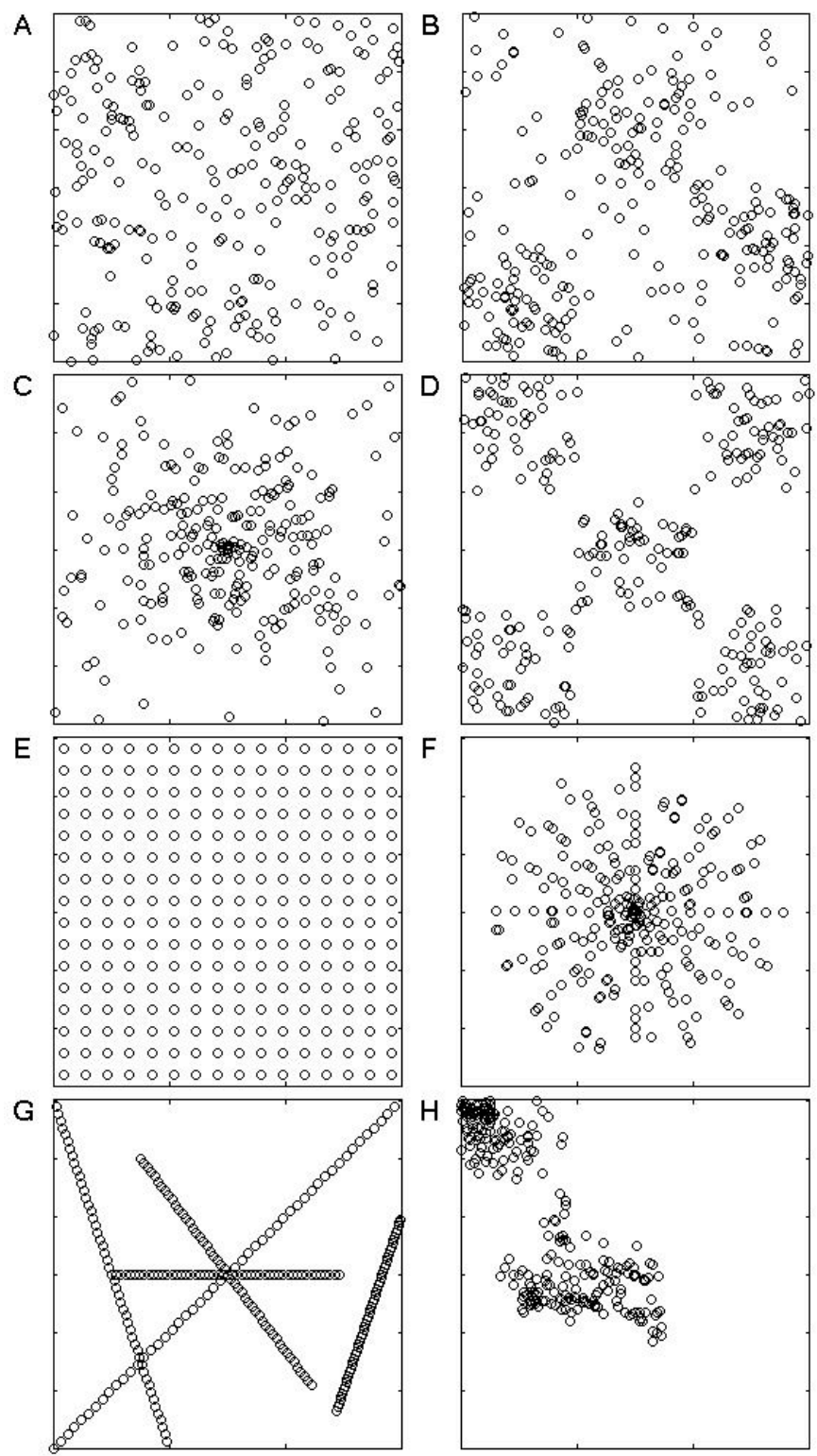

Figure 2.

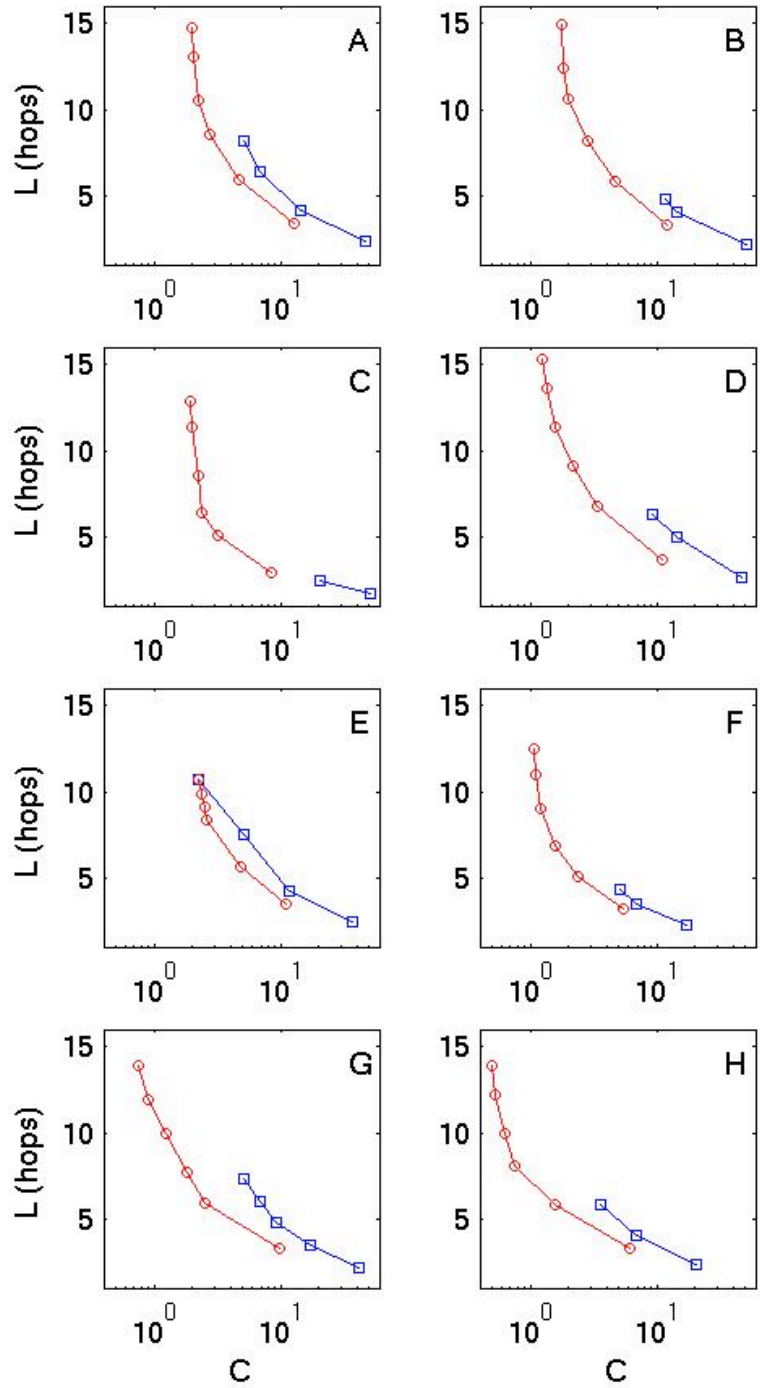

Figure 3.

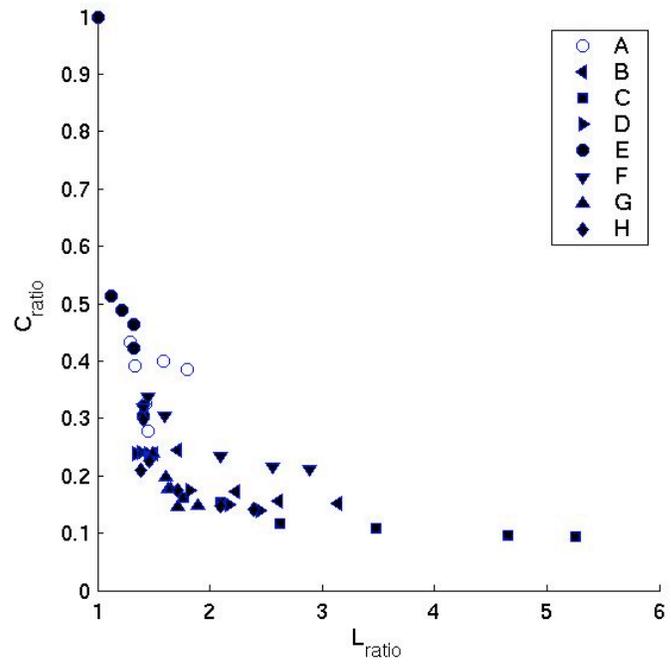

Figure 4.